\documentclass[%
 reprint,
 amsmath,amssymb,
 aps,
prb,
]{revtex4-1}

\usepackage{graphicx}
\usepackage{dcolumn}
\usepackage{bm}

\begin{document}

\preprint{APS/123-QED}

\title{Magnetic properties, domain wall creep motion and the Dzyaloshinskii-Moriya interaction in Pt/Co/Ir thin films}
 
\author{Philippa M. Shepley} 
\altaffiliation[Current address: ]{School of Chemical and Process Engineering, University of Leeds, Leeds,LS2 9JT, UK.}%
\author{Harry Tunnicliffe}%
\author{Kowsar Shahbazi}%
\author{Gavin Burnell}%
\author{Thomas A. Moore}%
\email{T.A.Moore@leeds.ac.uk} 
\affiliation{%
 School of Physics and Astronomy, University of Leeds, Leeds, LS2 9JT, UK.\\ 
}%

\date{\today}

\begin{abstract} 
We study the magnetic properties of perpendicularly magnetised Pt/Co/Ir thin films and investigate the domain wall creep method of determining the interfacial Dzyaloshinskii-Moriya (DM) interaction in ultra-thin films. Measurements of the Co layer thickness dependence of saturation magnetisation, perpendicular magnetic anisotropy, and symmetric and antisymmetric (i.e. DM) exchange energies in Pt/Co/Ir thin films have been made to determine the relationship between these properties. We discuss the measurement of the DM interaction by the expansion of a reverse domain in the domain wall creep regime. We show how the creep parameters behave as a function of in-plane bias field and discuss the effects of domain wall roughness on the measurement of the DM interaction by domain expansion. Whereas modifications to the creep law with DM field and in-plane bias fields have taken into account changes in the energy barrier scaling parameter $\alpha$, we find that both $\alpha$ and the velocity scaling parameter $v_{0}$ change as a function of in-plane bias field.

\end{abstract}

\pacs{Valid PACS appear here}
\maketitle


\section{\label{sec:intro}Introduction}

The magnetic behaviour of spin structures such as domain walls and skyrmionic bubbles in thin ferromagnetic films is determined by the interplay of three energy terms: magnetic anisotropy, Heisenberg exchange and the Dzyaloshinskii-Moriya (DM) interaction. The strengths of the symmetric and antisymmetric exchange interactions play a key role in determining the spin structure and energy of a domain wall, with the Heisenberg exchange favouring collinear alignment of spins and the DM interaction favouring orthogonal alignment of spins \cite{Moriya1960,Dzyaloshinsky1958,Heide2008,Yang2015,Nembach2015}. The magnetic anisotropy refers to the energetically favourable crystal axes or geometric directions that the magnetic moments align to. Here we study the balance of anisotropy and exchange energy terms by investigating how these properties are affected by magnetic layer thickness in Pt/Co/Ir thin films.

Domain walls in perpendicular magnetic anisotropy thin films were initially thought to form only in the Bloch structure, such that the energy of magnetic domain walls depended on the effective anisotropy constant $K_{\text{eff}}$ and the exchange stiffness $A$ as $\gamma= 4\sqrt{K_{\text{eff}}A}$. It is now understood that the interfacial DM interaction plays a role in the domain wall energy in perpendicular anisotropy thin films with broken inversion symmetry \cite{Thiaville2012,Je2013,Chen2013}. The DM interactions at Pt/Co and Ir/Co interfaces are generally held to be of opposite sign \cite{Hrabec2014,Yamamoto2017}, so are expected to contribute to a large net DM interaction when combined in an asymmetric trilayer such as Pt/Co/Ir. These trilayers are the building blocks of multilayers where skyrmionic structures have been detected \cite{Moreau-Luchaire}, making the understanding of their properties key to the development of devices based on the control and motion of chiral spin structures. 

When considering effects related to domain walls and related spin structures, it is important to consider all the contributions to the domain wall energy -- magnetic anisotropy, exchange stiffness and the DM interaction -- and how they interact. Using Kerr microscopy and SQUID-VSM, we show the effect of varying the thickness of the Co layer on the DM interaction and exchange stiffness in Pt/Co/Ir. We present a characterisation of the magnetic properties of Pt/Co/Ir thin films over a range of Co thicknesses exhibiting perpendicular anisotropy, which will be useful for ongoing studies on the physics of magnetic skyrmions and for designing devices from materials with exotic spin textures. 

The interfacial DM interaction has attracted much interest \cite{Lavrijsen2015,Khan2016,Moreau-Luchaire,Nembach2015,Petit2015,Vanatka2015,Hrabec2014,Chen2013} and has been investigated using different techniques\cite{Je2013,Vanatka2015,Nembach2015,Soucaille2016,LoConte2017,Zeissler2017}. A commonly used method, particularly for ultra-thin trilayer films with perpendicular magnetic anisotropy, is the expansion of reverse domains in the creep regime under in-plane bias fields, introduced by Je et al. \cite{Je2013}. This method has been applied in a range of cases, sometimes giving results that fit well to the modified creep model \cite{Je2013,Petit2015,Khan2016}, sometimes giving results that are more difficult to interpret \cite{Vanatka2015,Lavrijsen2015,Soucaille2016} or that do not give the same value for the DM energy as other methods \cite{Soucaille2016}. Where it fits well to experimental data and gives a clear result, the creep model can provide a value for the DM energy over a localised area (for example, close to a defect acting as a nucleation centre) in which other phenomena are being observed, or provide a lower limit for a thin film \cite{Zeissler2017}.

We measure the DM interaction by the method of Je et al. \cite{Je2013}, and use the results of our characterisation of Pt/Co/Ir films to discuss the limitations of the method and point a way towards further development of the technique and relevant creep theory. We show that the velocity scaling parameter, in addition to the energy barrier scaling parameter, changes as a function of applied in-plane bias field. We build on recent work on the creep motion of domain walls \cite{Jeudy2016} to analyse the changes to the creep parameters. We describe how domain wall roughness has an effect on how well the modified creep equation \cite{Je2013} models the domain wall velocity when high in-plane bias fields are applied. For measurements with low bias fields, in thin films with low magnetic roughness and without strong pinning points and with low interfacial DM interaction, the modified creep model can give a measure of the DM field. Advances in the relevant creep theory have the potential to expand situations where the domain wall creep technique can be used for DM measurements.

\section{\label{sec:exp}Experimental methods} 
 
We study the magnetic properties of a series of thin films of Ta(4.5nm)/Pt(4nm)/Co(t)/Ir(5nm), with Co thickness $t$ varying between 0.56 nm and 1.1 nm, deposited onto thin glass substrates by dc magnetron sputtering \cite{Shepley2015}. The magnetic properties were measured by a combination of SQUID-VSM magnetometry and magneto-optical Kerr effect (MOKE) microscopy. The magnetic anisotropy field $H_{\text{k}}$ was measured both from in-plane SQUID-VSM hysteresis loops and from polar MOKE versus in-plane field moment rotation (similar to the method used previously \cite{Shepley2015}, with the change in the polar MOKE signal being proportional to the change in the out-of-plane magnetisation component). The results from the two methods are consistent, with the values from the Kerr method used here, since this measurement technique is local to the region of the film where the domain wall velocities are measured. The saturation magnetisation $M_{\textbf{s}}$ was recorded from SQUID-VSM hysteresis loops, and the exchange stiffness $A$ was found by fitting a Bloch $T^{3/2}$ law to normalised SQUID-VSM moment versus temperature curves \cite{Talagala2002,Nembach2015}.  
 
Domain wall velocities were measured by quasi-static domain wall imaging using a wide-field MOKE microscope. A field pulse was applied to nucleate a reverse domain, an image was recorded, then a second pulse was applied to move the domain wall and a second image was recorded. The difference of the two images shows a bright region through which the domain wall has moved and, knowing the duration of the field pulse, the domain wall velocity can be calculated. The DM field was estimated from the asymmetric expansion of reverse domains, imaged by MOKE microscopy, under out-of-plane driving fields and applied in-plane bias fields \cite{Je2013}. The minima of velocity versus $H_{\text{x}}$ curves occur when the DM field is balanced by the applied in-plane bias field. We examine the validity of this method for determining the DMI by comparing the measured velocity vs $H_{\text{x}}$ curves to the modified creep law proposed by Je et al.\cite{Je2013}.

\begin{figure} 
\centering 
\includegraphics[width=8cm]{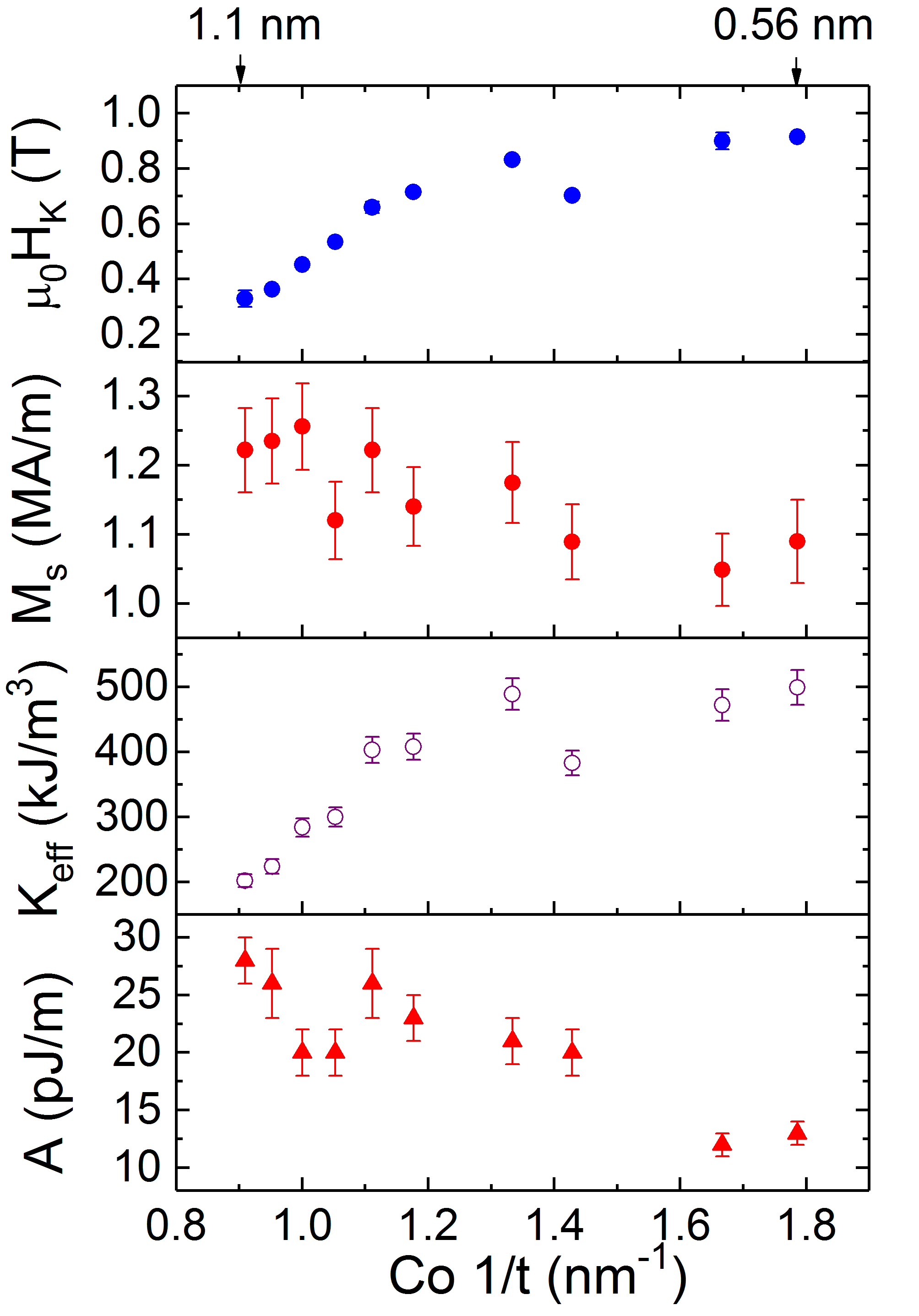} 
\caption{Measured values of anisotropy field, saturation magnetisation, effective anisotropy constant and exchange stiffness are plotted for Pt/Co(t)/Ir for Co thicknesses of t = 0.56 - 1.1 nm.} 
\label{fig_4panel_Hk_Ms_K_A} 
\end{figure} 
 
\section{\label{sec:magprop}Results and discussion} 
 
\subsection{\label{sec:}Magnetic characterisation} 

The magnetic properties of Pt/Co/Ir thin films depend on the thickness of the Co layer. Figure \ref{fig_4panel_Hk_Ms_K_A} shows how the anisotropy field $H_{\text{K}}$, saturation magnetisation $M_{\text{s}}$, effective anisotropy constant $K_{\text{eff}}$ and exchange stiffness $A$ vary with Co thickness. The effective anisotropy is calculated using the measured $H_{\text{K}}$ and $M_{\text{s}}$ as $K_{\text{eff}}=\frac{1}{2}\mu_{0}H_{\text{K}}M_{\text{s}}$. The size of the saturation magnetisation $M_{\text{s}}$ and the exchange stiffness $A$ decrease with $1/t$ from  $M_{\text{s}}$ = 1.25 MA/m for the thickest film with $t$ = 1.1 nm to $M_{\text{s}}$ = 1.05 MA/m for the thinnest film with $t$ = 0.56 nm, and $A$ = 28 pJ/m for $t$ = 1.1 nm reducing to 12 pJ/m at $t$ = 0.56 nm. The magnitudes and observed trend of exchange stiffness $A$ with thickness are similar to calculated values for Pt/Co/Pt thin films \cite{Metaxas2007}. The exchange stiffness exhibits a similar behaviour with ferromagnetic layer thickness as reported by Nembach et al. \cite{Nembach2015} in a Ni$_{80}$Fe$_{20}$/Pt system. The effective anisotropy constant is between 400 and 500 kJ/m$^{3}$ for most of the Co thickness range, but decreases rapidly as the thickness increases above 0.95 nm.  

\subsection{\label{sec:DM}Dzyaloshinskii-Moriya energy} 

The DM field in Pt/Co/Ir was measured by expanding a reverse domain in the creep regime, as first proposed by Je et al \cite{Je2013} and subsequently employed by a number of investigators for a range of magnetic thin films \cite{Hrabec2014,Petit2015,Lavrijsen2015,Vanatka2015,Khan2016,Lau2016}.  The model describing the domain expansion is a version of the creep law with the energy barrier scaling parameter modified to include the change to the domain wall energy due to an applied in-plane bias field. The domain wall creep velocity driven by an out-of-plane magnetic field $H_{\text{z}}$ is given by 
\begin{equation} 
v=v_{0}\exp\Big[-\alpha(\mu_{0}H_{z})^{-\frac{1}{4}}\Big],
\label{eq_creep} 
\end{equation}where $v_{0}$ is a velocity scaling parameter and the energy barrier scaling parameter $\alpha$ can be written as  
\begin{equation} 
\alpha=\alpha _{0} \bigg(\frac{\gamma(H_{x})}{\gamma(0)}\bigg)^{\frac{1}{4}}=\frac{T_{dep}}{T}\big(\mu_{0}H_{dep}\big)^{\frac{1}{4}} \bigg(\frac{\gamma(H_{x})}{\gamma(0)}\bigg)^{\frac{1}{4}},
\label{eq_alpha} 
\end{equation} 
where $\alpha _{0}$ depends on the pinning energy with no in-plane bias field $T_{dep}/T$ and the depinning field $H_{\text{dep}}$, as well as the domain wall energy in an applied in-plane field, $\gamma(H_{\text{x}})$. Note that $\alpha = \alpha _{0}$ when no in-plane bias field is applied. The field-dependent domain wall energy term  includes the DM field and is different depending on whether the domain wall is truly in the N\'{e}el configuration \begin{equation} 
\gamma_{N}=\gamma + 2K_{D}\delta - \pi \delta \mu_{0}M_{s} |H_{x} + H_{DM}|, 
\label{eq_neel} 
\end{equation} 
or retains some Bloch character 
\begin{equation} 
\gamma_{BN}=\gamma - \frac{ \delta(\pi\mu_{0}M_{s})^{2}}{8K_{D}} (H_{x} + H_{DM})^{2}. 
\label{eq_BN} 
\end{equation} 
The energy of a pure Bloch wall is $\gamma =4\sqrt{AK_{eff}}$,  $\delta =\sqrt{A/K_{eff}}$ relates to the wall width, and the domain wall shape anisotropy \cite{Tarasenko1998,Thiaville2012} is $K_{D}=2\ln(2)t\mu_{0}M_{s}^{2}/\pi\delta$. The wall becomes N\'{e}el when the DM and bias fields are sufficient to overcome the wall shape anisotropy, $\mu_{0}|H_{x} +H_{DM}| < 4K_{D}/\pi M_{s}$. 
  
\begin{figure} 
\centering 
\includegraphics[width=8cm]{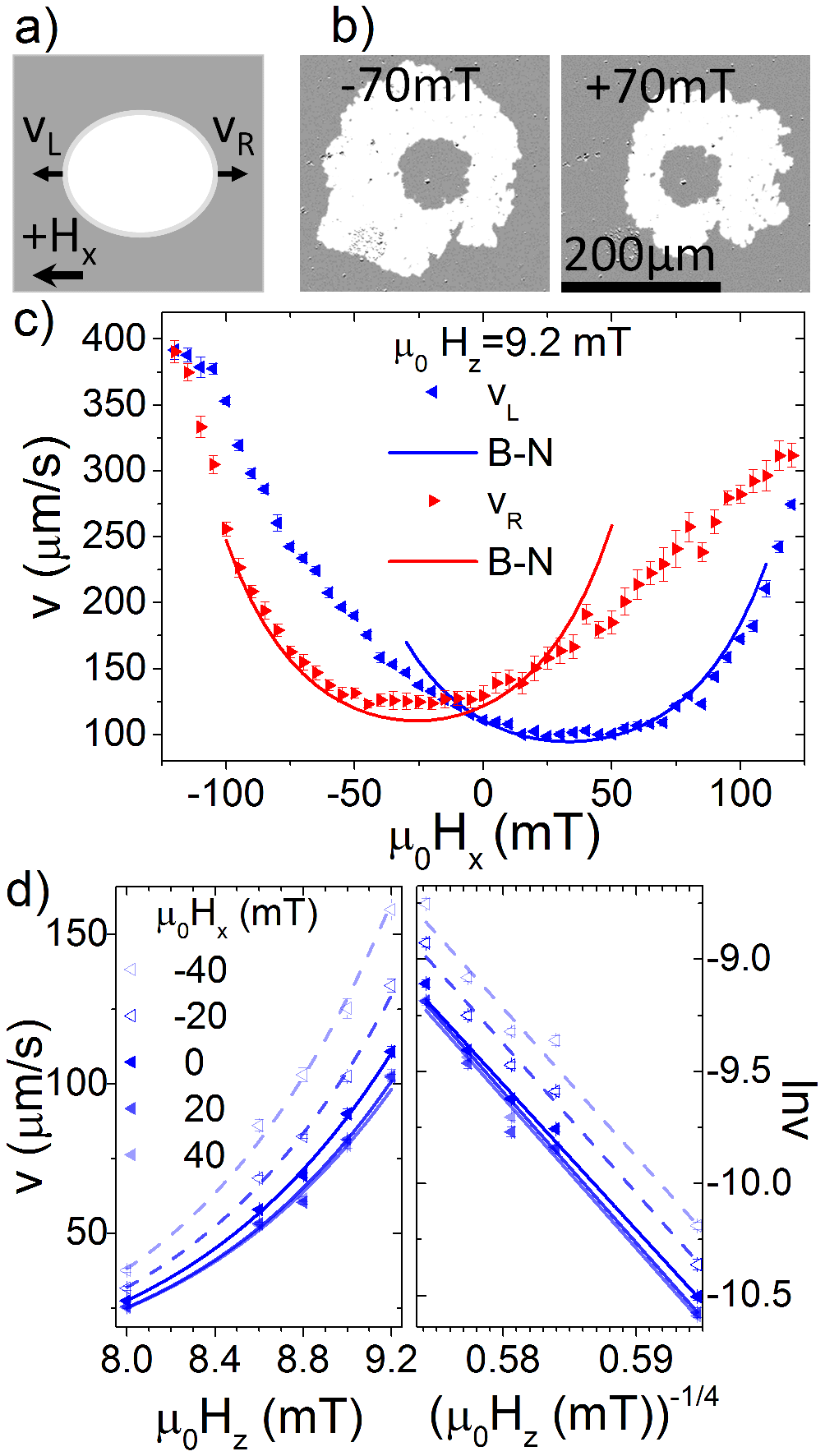} 
\caption{a) Schematic of a domain expanding under an applied out-of-plane field $H_{\text{z}}$ with an in-plane bias field $H_{\text{x}}$ and b) examples of domain expansion difference images. The dark area in the centre is the initial domain and the bright area is the region that reverses due to motion of the wall driven by $H_{\text{z}}$. c) The left and right pointing triangles are the velocities of domain walls on the left and right sides of the domain in Pt/Co(0.7nm)/Ir plotted against the bias field $H_{\text{x}}$, under a driving field of $\mu_{0}H_{\text{z}}$ = 9.2 mT. The lines show the velocity curves for Bloch-N\'{e}el walls calculated from the modified creep model using creep parameters measured at $H_{\text{x}}$=0 and magnetic properties shown in Figure \ref{fig_4panel_Hk_Ms_K_A}. d) Creep velocity $v$ plotted against the out-of-plane driving field for applied bias fields, including $\mu _{0} H _{x} = $ 0 $mT$, for the left side of the domain in  Pt/Co(0.7nm)/Ir. The natural logarithm of velocity for the same data is plotted against $(\mu _{0} H _{z})^{-1/4}$. The lines are fits of the data to the creep law given in Equation \ref{eq_creep} and the linear form using the natural logarithm of velocity given in Equation \ref{eq_lnv}.} 
\label{fig_creep_vs_Hx} 
\end{figure} 

There has been some variation in the success of this model in descibing the shape of velocity versus $H_{\text{x}}$ curves. The model was shown to work well for the Pt/Co/Pt films studied by Je et al. \cite{Je2013}, and also well for the Ta/CoFeB/MgO studied by Khan et al. \cite{Khan2016}. Lavrijsen et al. \cite{Lavrijsen2015} find a wide variety of $v(H_{\text{x}})$ curves for Pt/Co/Pt that are not symmetric around the minimum velocity. Vanatka et al. \cite{Vanatka2015} show in Pt/Co/Gd films that velocity versus $H_{\text{x}}$ curves in the flow regime can give curves that are symmetric around the velocity minimum in cases where the creep regime yields unclear results. However, since the flow regime is not always easily accessible due to the high fields required or the onset of Walker breakdown, where possible, the creep regime technique may be the most convenient.

The velocity versus $H_{\text{x}}$ curves we measure for the Pt/Co/Ir thin films are close to the shape expected from the modified creep law, with some variations. Figures \ref{fig_creep_vs_Hx} a and b give a schematic of the expansion of a domain, defining the field and velocity directions, and examples of expanding domains in Pt/Co(0.7nm)/Ir under positive and negative in-plane fields. A representative example of velocity vs $H_{\text{x}}$ curves extracted from MOKE images for right and left moving walls (Figure \ref{fig_creep_vs_Hx}c) shows that the curves have clear minima and, close to the minima, are approximately symmetric around the lowest velocity values. 

We have extracted the values of $\alpha$ and $\ln v_{0}$ measured at $H_{\text{x}}$=0 by linear least squares fits of the creep law (Equation \ref{eq_creep}) in the natural log form \cite{Shepley2015} 
\begin{equation}
\ln v=\ln v_{0}-\alpha (\mu_{0}H_{z})^{-\frac{1}{4}},
\label{eq_lnv}
\end{equation}
to plots of $\ln v$ versus $(\mu_{0}$H$_{z})^{-\frac{1}{4}}$. Creep velocity data measured at different out-of-plane driving fields are shown in Figure \ref{fig_creep_vs_Hx}d. Using the values of $\alpha$ and $\ln v_{0}$ extracted at $H_{x}=$ 0, the measured magnetic properties shown in Figure \ref{fig_4panel_Hk_Ms_K_A}, and the DM field taken from the minimum of the curve we have calculated velocity versus $H_{\text{x}}$ curves expected from the modified creep model (Equation \ref{eq_creep}). The calculated curves match quite well to the data, suggesting that the field at which the velocity is a minimum is a reasonable estimate of the DM field.

\begin{figure} 
\centering 
\includegraphics[width=8cm]{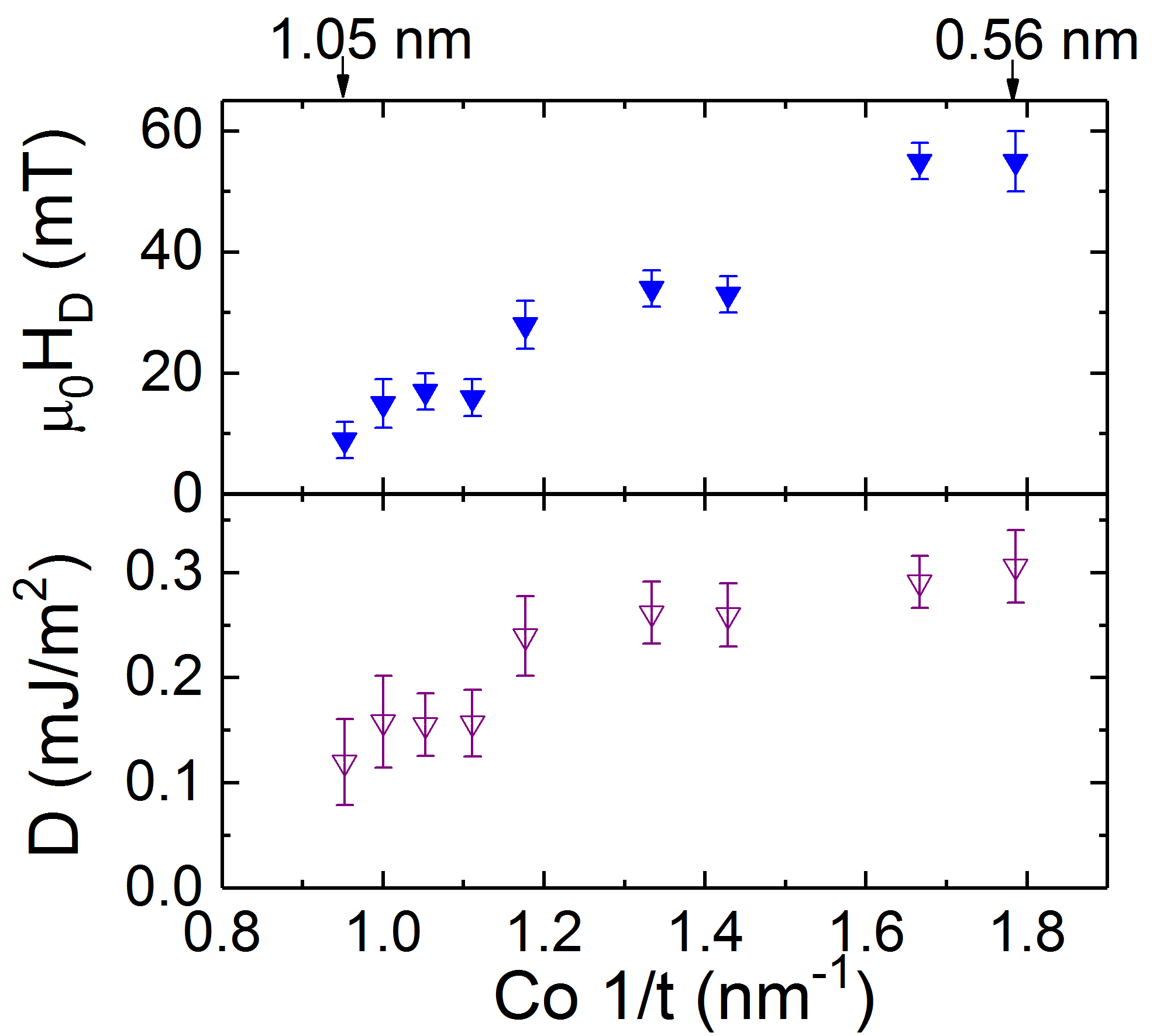} 
\caption{Measured values of DM field and DM energy density are plotted for Pt/Co(t)/Ir for Co thicknesses of t = 0.56 - 1.05 nm.} 
\label{fig_2panel_Hd_D} 
\end{figure} 

Taking the minima of the velocity versus $H_{\text{x}}$ curves as the DM field gives values that depend on the Co thickness. Figure \ref{fig_2panel_Hd_D} shows the measured values of DM field, given as $\mu_{0}H_{\text{D}}$ and the DM energy density D, where D is calculated from $D=\mu_{0}H_{\text{D}}M_{\text{s}}\delta$. The DM field is smaller for Pt/Co/Ir films with thicker Co and larger for those with thinner Co. None of the Pt/Co/Ir films have a sufficiently large DM field to fully overcome the domain wall shape anisotropy that favours Bloch walls, so all will have domain walls with a combination of the Bloch and left-handed N\'{e}el spin structures. 

The reduction of the DM field with increasing Co thickness is consistent with the understanding of the DM interaction in heavy-metal/ferromagnet trilayers as an interfacial effect.  The trend with Co thickness shows that we might expect there to be no net DM field in Pt/Co/Ir films with Co layers thicker than 3.8 nm, at which point the dominance of in-plane volume magnetic anisotropy indicates that the magnetic properties have become less dependent on the interfaces \cite{Shepley2015}.
Interfacial effects are generally expected to scale as $1/t$. In some other studies that have investigated the dependence of $D$ on magnetic layer thickness it has been possible to fit a straight line of the form $D \propto 1/t$ to DM energy data for thin film systems \cite{Kim2016a,Han2016}. Other studies show a linear dependence but with a non-zero intercept, suggesting some contribution from atoms that do not lie at a sharp heavy metal/ferromagnet interface layer \cite{Belmeguenai2015}. A similar result to ours was obtained by Nembach et al for NiFe/Pt, where the measured DM interaction did not exhibit a strict $1/t$ dependence \cite{Nembach2015}.

\begin{figure}[ht!]
\centering 
\includegraphics[width=8cm]{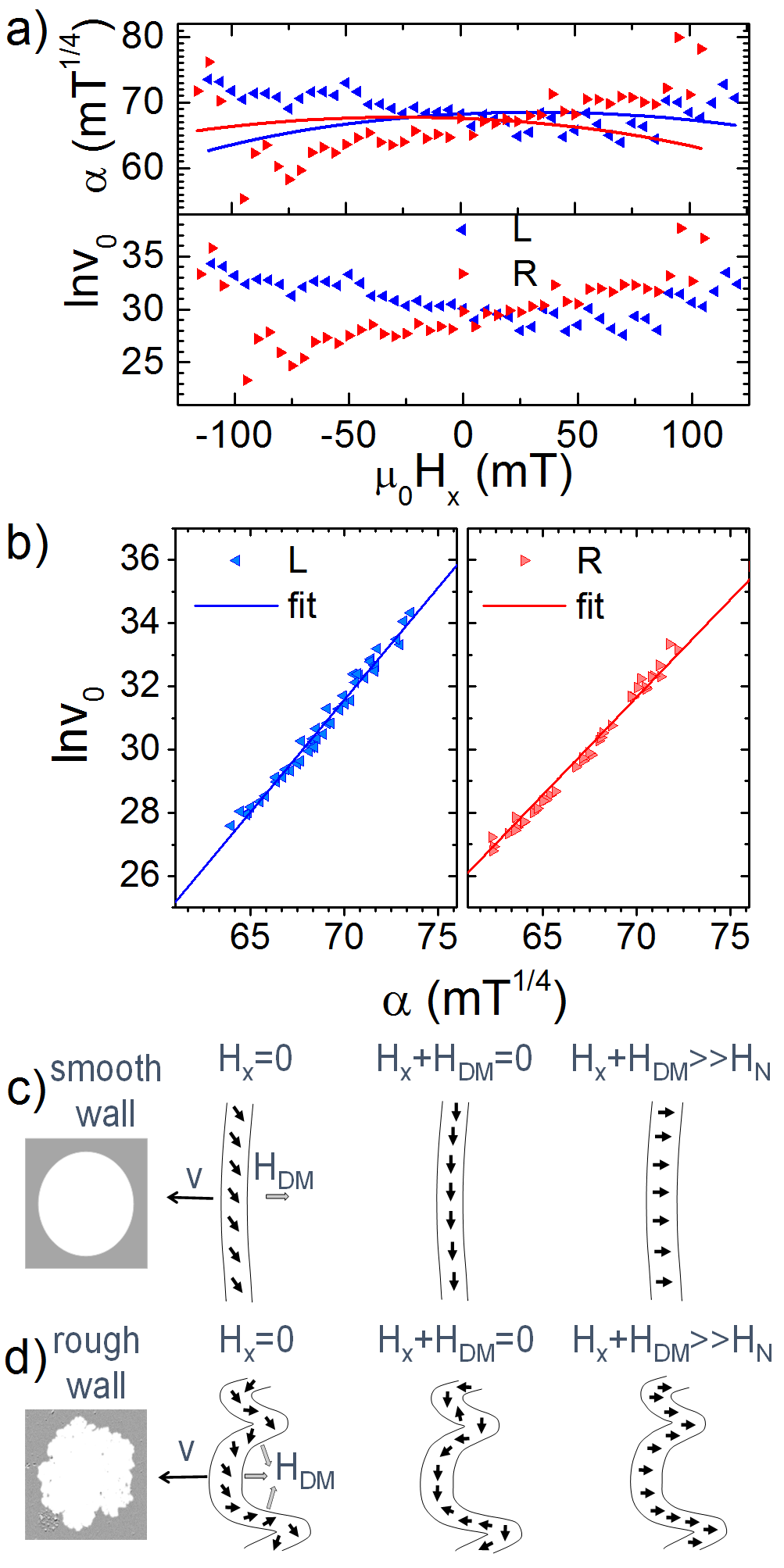} 
\caption{a) The creep parameters extracted from velocity versus $H_{\text{z}}$ at different values of $H_{\text{x}}$ are plotted against the applied bias field and b) plotted against each other. The blue and red data are for the left and right moving domain walls. The solid lines in the topmost panel of a) show the energy barrier scaling parameter $\alpha$ calculated from the parameters measured at $H_{\text{x}}$ = 0, in b) the lines are fits to Equation \ref{Eq_creep_TM1}. The errors on the data are typically $\pm$2mT$^{1/4}$ in $\alpha$ and $\pm$1 in $\ln v_{0}$, and are not plotted so that the data points are clearly visible.  c) and d) present sketches showing the difference in the magnetisation direction at the centre of perfectly smooth and rough domain walls under applied bias field $H_{\text{x}}$.} 
\label{fig_alpha_lnv0_rough} 
\end{figure}
 
We note that in our case the measured DM field is due to an ultrathin Co layer with both a Pt/Co and a Co/Ir interface, which give rise to different magnitudes and chiralities, and it cannot be assumed that the DM field measured in ultrathin Co layers will extrapolate to zero for very thick layers ($1/t \rightarrow$ 0).  Making such an assumption would imply that the DM interaction is relevant only to the first plane of atoms at a sharp interface, which is not realistic \cite{Yang2015}. 
Theoretical work by Yang et al. \cite{Yang2015} has shown that the DM energy at an ideal Pt/Co interface does not vary linearly with $1/t$. Yang et al show that the DM energy comes mainly from spin orbit coupling interactions with neighbouring atoms. Their calculations show that while most of the contribution to the DM interaction comes from the heavy metal/ferromagnet interface, with the DM contribution concentrated mainly in the first magnetic layer, there is also a small DM contribution from the other atomic layers. The contribution from the non-interface layers of ferromagnetic atoms has an opposite sign to the interface contribution. Since there is a small contribution to $D$ from Co atoms in our films that are not in direct contact with Pt or Ir, we do not expect to find an exact $D \propto 1/t$ form for the measured DM energy.

The thickness dependence of the DM field and anisotropy are similar, however, there are some differences such as a clear change in the slope of the anisotropy versus $1/t$ that is not as apparent in the DM field. While both properties are dependent on the interfaces, the dependence is not the same, which is consistent with the finding from calculations by Yang et al. that $D$ is not correlated to an enhancement in the Pt moment that contributes to the perpendicular anisotropy \cite{Nakajima1998,Yang2015}.

\subsection{\label{sec:DMrough}Dzyaloshinskii-Moriya interaction and domain wall creep motion}

We can further investigate the velocity versus $H_{\text{x}}$ creep motion by extracting creep parameters \cite{Lavrijsen2015}. The velocity versus $H_{\text{x}}$ curves shown in Figure \ref{fig_creep_vs_Hx} were taken at five different out-of-plane fields so that the creep parameters $v_{0}$ and $\alpha$ could be extracted from linear least squares fits of the natural log form of the creep law (Equation \ref{eq_lnv}) to plots of $\ln v$ versus $(\mu_{0}$H$_{z})^{-\frac{1}{4}}$. Examples of the $\ln v$ versus $(\mu_{0}$H$_{z})^{-\frac{1}{4}}$ with linear fits are shown in Figure \ref{fig_creep_vs_Hx}d. The results of the fitting, represented by the data points in Figure \ref{fig_alpha_lnv0_rough}a show that both the energy barrier scaling parameter $\alpha$ and the velocity scaling parameter $v_{0}$ change with applied in-plane field. The variation of $\alpha$ with $H_{\text{x}}$ expected from the modified creep model, calculated using the measured magnetic properties of Pt/Co/Ir and plotted as solid lines in Figure \ref{fig_alpha_lnv0_rough}a, does not have the same shape as the data. A limitation of the modified creep law that becomes clear from analysis of the creep parameters is the omission of a dependence of $\ln v_{0}$ on $H_{\text{x}}$. While the modified creep model includes a variation in the energy barrier scaling parameter $\alpha$ with $H_{\text{x}}$, it does not account for the changes in $\ln v _{0}$, which can be seen from Figure \ref{fig_alpha_lnv0_rough}a to have a dependence on $H_{\text{x}}$ very similar to that of $\alpha$.  
 
We can look to recent work on the creep law to see how we might expect $\ln v_{0}$ and $\alpha$ to have similar $H_{\text{x}}$ dependence. Jeudy et al. \cite{Jeudy2016} showed that 
\begin{equation} 
v=v_{0}'(H_{dep},T)\exp \Big[ - \frac{\Delta E}{kT}\Big],
\label{Eq_creep_juedy1}
\end{equation}
where $v_{0}'(H_{dep},T)$ is a velocity scaling parameter that can vary as a function of depinning field and temperature, T is the temperature and the universal creep energy barrier scaling parameter is 
\begin{equation} 
\Delta E =k T_{dep} \Big[ \Big( \frac{H_{z}}{H_{dep}}\Big)^{-\frac{1}{4}}-1\Big],
\label{Eq_creep_juedy2}
\end{equation}
gives good agreement with creep velocity data from a variety of materials and a large driving field range. For the purpose of investigating the extracted creep parameters we may write the creep law as 
\begin{equation} 
\ln v = \frac{T_{dep}}{T}+\ln v_{0}'(H_{dep},T)
- \frac{T_{dep}}{T} (\mu_{0}H_{dep})^{\frac{1}{4}}(\mu_{0}H_{z})^{-\frac{1}{4}}.
\label{Eq_creep_juedy3}
\end{equation}
As shown in Equation \ref{eq_alpha}, the creep energy scaling factor is given by 
\begin{equation}
\alpha =\frac{T_{dep}}{T} (\mu_{0}H_{dep})^{\frac{1}{4}}.
\label{Eq_creep_juedy5}
\end{equation}
Since $\alpha$ is the gradient extracted from our straight line fits of creep velocity measurements under in-plane bias field, the intercept that we call the velocity scaling parameter $\ln v_{0}$ is then given by 
\begin{equation}
\ln v_{0} = \frac{T_{dep}}{T}+\ln v_{0}'(H_{dep},T).
\label{Eq_creep_juedy4}
\end{equation}

The extracted velocity scaling parameter can be written as a function of the energy scaling parameter:
\begin{equation}
\ln v_{0} = \frac{\alpha}{(\mu_{0}H_{dep})^{\frac{1}{4}}}+\ln v_{0}'(H_{dep},T).
\label{Eq_creep_TM1}
\end{equation}
When we apply an in-plane bias field $H_{\text{x}}$, the energy scaling parameter $\alpha$ depends on $[\gamma(H_{\text{x}})]^{\frac{1}{4}}$, so the extracted creep parameters $\alpha$ and $\ln v_{0}$ can be expected to have a similar dependence on $H_{\text{x}}$. 

In Figure \ref{fig_alpha_lnv0_rough}b we plot the creep parameter data as $\ln v_{0}$ vs $\alpha$ and show fits to Equation \ref{Eq_creep_TM1}. Under the assumption that the depinning field $H_{dep}$ doesn't change when an in-plane bias field is applied, Equation \ref{Eq_creep_TM1} will take the form of a straight line. The straight lines fitted to the plots of $\ln v_{0}$ vs $\alpha$ in Figure \ref{fig_alpha_lnv0_rough}b confirm that this model gives good agreement with the data, and demonstrates that both $\ln v_{0}$ and $\alpha$ can be expected to change as a function of in-plane bias field. 

In studying the difference between the modified $v(H_{\text{x}})$ creep model and the data, the roughness of the wall is also an important factor since the modified creep model must assume a smooth domain wall. Figure \ref{fig_alpha_lnv0_rough}c and \ref{fig_alpha_lnv0_rough}d show the difference we might expect between the behaviour of a smooth and rough Bloch-N\'{e}el (BN) wall under applied in-plane bias fields. The DM field always acts perpendicular to the line of the wall, so for a smooth wall, the DM field and applied bias field in the section of the wall that we measure are always parallel or anti-parallel, and the magnetisation direction in the centre of the wall will rotate due to the relative size of the two fields. In a rough wall, the direction of the DM field will vary with respect to the applied bias field, so the balance between the two fields will vary along the wall and the magnetisation direction in the centre of the wall will only behave as we expect when for sections of wall where the DM and bias field align along a common axis. As a consequence of the variation in angle between the two fields, a rough wall can never take on a fully Bloch or fully N\'{e}el structure when a bias field is applied. This has the effect of suppressing changes in velocity due to the variation in wall energy with bias field and preventing the wall from reaching the higher velocities that should be possible for a N\'{e}el wall, as can be seen in Figure \ref{fig_creep_vs_Hx}.

\section{\label{sec:summary}Summary} 
 
We have measured the magnetic properties of Pt/Co/Ir thin films, including the perpendicular magnetic anisotropy energy, Heisenberg exchange stiffness and interfacial DM energy, with respect to varying Co layer thickness. We have characterised the magnetic properties on which the energies of spin structures such as domains wall depend, with respect to varying Co layer thickness. 


We have investigated how the creep parameters behave as a function of in-plane bias field during DM energy measurments. We used a domain wall creep $v(H_{\text{x}})$ method to estimate the strength of the DM interaction and found that the creep parameters $\ln v_{0}$ and $\alpha$ do not behave as expected, particularly at large in-plane bias fields. The modified creep model proposed by Je et al. \cite{Je2013} takes into account the change in the creep energy scaling parameter $\alpha$ under applied in-plane bias fields. We find that both the energy barrier scaling parameter $\alpha$ and the velocity scaling parameter $\ln v_{0}$ change when an in-plane bias field is applied. We can also understand some of the differences between the data and modified creep model in Figure \ref{fig_2panel_Hd_D} as relating to the roughness of the domain walls. Wall roughness causes a variation in the angles between the bias field and DM field along the wall, reducing the effect of the bias field on the wall velocity and slowing the wall at large $H_{\text{x}}$ values.

The characterisation of Pt/Co/Ir trilayers over a range of Co thicknesses will be useful for development of materials for devices based on domain walls or other spin structures. We have demonstrated the limitations of the creep-method measurements of interfacial DM energy, used current understanding of the creep law to investigate how the creep parameters are affected by in-plane bias fields, and described the role that domain wall roughness plays when a bias field is applied. We hope this work will lead to further developments in the theory of creep motion of magnetic domain walls in perpendicular systems with in-plane bias fields.

\section*{ACKNOWLEDGMENT}
The authors acknowledge financial support from EPSRC (Grants No. EP/K003127/1 and No. EP/M000923/1).

Data associated with this paper are available from https://doi.org/10.5518/353.

\bibliographystyle{IEEEtran}
\bibliography{shep_refs}
\end{document}